\documentclass[aps,prb,twocolumn,groupedaddress]{revtex4}
\usepackage{epsfig}
\usepackage[american]{babel}
\usepackage{amsmath}
\usepackage[dvips]{color}

\usepackage[latin2]{inputenc}
\usepackage[T1]{fontenc}
\usepackage[american]{babel}
\usepackage{color}
\usepackage{graphicx}
\usepackage{graphics}

\newcommand{\Ei}{\operatorname{Ei}}

\newcommand{\RS}{Rayleigh-Schr\"{o}dinger}

\begin{document}

\title{Holstein light quantum polarons on the one-dimensional lattice}

\author{O. S. Bari\v si\' c}

\email{obarisic@ifs.hr}

\affiliation{Institute of Physics, Bijeni\v cka c. 46, HR-10000 Zagreb, Croatia}

\begin{abstract}

The polaron formation is investigated in the intermediate regime of the Holstein model by using an exact diagonalization technique for the one-dimensional infinite lattice. The numerical results for the electron and phonon propagators are compared with the nonadiabatic weak- and strong-coupling perturbation theories, as well as with the harmonic adiabatic approximation. A qualitative explanation of the crossover regime between the self-trapped and free-particle-like behaviors, not well-understood previously, is proposed. It is shown that a fine balance of nonadiabatic and adiabatic contributions determines the motion of small polarons, making them light. A comprehensive analysis of spatially and temporally resolved low-frequency lattice correlations that characterize the translationally invariant polaron states is derived. Various behaviors of the polaronic deformation field, ranging from classical adiabatic for strong couplings to quantum nonadiabatic for weak couplings, are discussed.

\end{abstract}

\maketitle

\section{Introduction}

Charge transport observed in the limit of the low electronic density is characterized by significant polaronic effects when the electron-phonon coupling is strong enough. Indeed, the effects of this kind have been reported recently in a number of experiments on systems ranging from colossal magnetoresistive manganites to high-$T_c$ superconductors,\cite{Edwards,Devreese,Mannella} which are currently the subject of great scientific interest. In this context, a better understanding of the polaron states is of particular importance. The local polaronic correlations and the polaron dispersion properties are characterized by adiabatic and nonadiabatic contributions combined with the effects due to the discreteness of the lattice. In this respect the polaron problem contains most of the conceptual elements important for the dynamics of other problems with localized solutions. In particular, the theory of quantum polarons has its analogies in the more general context of soliton physics, invoked for broad classes of materials.

The experimental evidence of strongly coupled polaronic carriers is frequently provided in terms of the thermally activated d.c. conductivity, which is accompanied by a broad contribution to the mid-infrared optical conductivity. \cite{Wang,Emin} On the other hand, in photoemission spectroscopy measurements, one observes a small electron quasi-particle weight at the Fermi level, with most of the spectral weight transferred towards high frequencies.\cite{Perfetti,Alexandrov} In order to rationalize such strong polaronic correlations, assuming a local electron-phonon interaction, one may apply the Holstein model. Within the Holstein model,\cite{Holstein} in the strong-coupling limit the electron-phonon spatial correlations reduce to a single lattice site and one obtains heavy, strongly-pinned, small polaron states characterized by a large lattice distortion. For such states the polaron binding energy $\varepsilon_p$ determines the thermal activation energy, $\omega_{d.c.}\approx\varepsilon_p/2$, and the optical activation energy $\omega_{opt}\approx2\varepsilon_p$.\cite{Firsov} Yet, even for quite strong couplings the adiabatic spreading of the lattice distortion over a few lattice sites might significantly reduce the pinning barrier for the polaron motion, $\omega_{d.c.}\ll\omega_{opt}$. In particular, in the one-dimensional Holstein model weakly-pinned small polarons with a considerable binding energy are obtained between the strong- and weak-coupling limits for moderate values of the adiabatic parameter. One of the key objectives of this work is to achieve a qualitative understanding of this interesting parameter regime, which can be relevant for unconventional small-polaron behaviors reported for various materials.

The accurate treatment of adiabatic correlations is a difficult problem for itinerant polarons. Several numerical methods have been recently suggested for the one-dimensional Holstein model in order to spectrally resolve the polaronic correlations in the single-electron and optical conductivity spectra.\cite{Wellein,Hohenadler,Filippis} These studies clearly show the fundamental difference between the nonadiabatic and adiabatic regimes, although the reliability of the methods at low frequencies is not entirely clear. In the adiabatic regime one should obtain the excited coherent polaron bands with finite spectral weight below the phonon threshold, each of the bands corresponding to one of the adiabatically softened local phonon modes.\cite{Barisic2} Yet, the spectral weight assigned to such low-frequency adiabatic correlations seems to be reported in the literature only for the case of Diagrammatic Quantum Monte Carlo calculations. \cite{Mishchenko2} The Dynamic mean-field theory (DMFT) applied to the Holstein polaron problem,\cite{Ciuchi2} exact in the infinite dimensional limit, gives only one excited coherent polaron band below the phonon threshold. For low-dimensional systems the approaches\cite{Sumi,Bronold} like DMFT that treat the electron self-energy as a local quantity, $\Sigma(k, \omega)\rightarrow\Sigma(\omega)$, provide approximate results. Namely, in order to describe correctly the local adiabatic correlations within the diagrammatic approach, the vertex corrections involving phonons at different lattice sites $i\neq j$, which contribute to the $k$-dependent (nonlocal) part of the electron self-energy $\Sigma(k,\omega)$, have to be taken into account. In fact, the important vertex corrections turn out to be those within the range of the adiabatic correlation length.\cite{Barisic4} This length is restricted by the distance over which the electron can move before the lattice distortion evolves significantly in time.

In order to obtain the possibly most accurate picture of the low-frequency polaronic correlations a previously developed approach\cite{Bonca1,Barisic2} of exact diagonalization of the infinite lattice polaron problem is employed here. By providing eigenenergies and corresponding wave functions, this approach, unlike any other developed up to now, gives the complete information about the coherent part of the polaron spectrum below the phonon threshold. In addition to the polaron spectra and electron spectral properties, the phonon spectra that describe the displacement and momentum correlations of the lattice can be analyzed in this way. In fact, this work provides, to our knowledge, the first comprehensive study of lattice properties within the translationally invariant polaron theory. According to the regime of parameters, qualitatively different behaviors of the lattice deformation field are obtained in the low-frequency range. Such findings can be of experimental interest, for example, in the context of recent neutron diffuse scattering (Huang scattering) or pulsed neutron-diffraction measurements.\cite{Louca}

\section{General\label{SecGeneral}}

The Holstein polaron arises from the local interaction of the electron in the nearest-neighbor tight-binding band with the dispersionless branch of optical phonons. For the one-dimensional lattice the Hamiltonian reads

\begin{eqnarray}
\hat{H}&=&-2t\sum_k\cos(k)\;c_k^{\dagger}c_k+
\omega_0\sum_qb^\dagger_qb_q\nonumber\\&-&
g/\sqrt N\sum_{k,q}c_{k+q}^\dagger c_k\;(b^{\dagger}_{-q}+b_q)
\label{HolHam}\;,
\end{eqnarray}

\noindent where $N$ is the number of lattice sites ($N\rightarrow\infty$). $c_k^{\dagger}$ ($c_k$) and $b^{\dagger}_q$ ($b_q$) are the creation (annihilation) operators for the electron state with momentum $k$ and phonon state with the momentum $q$, respectively. Besides $t/\omega_0$ and $g/\omega_0$, the two other parameters used here are the binding energy of the polaron in the atomic limit ($t=0$), $\varepsilon_p=g^2/\omega_0$, and the dimensioneless parameter $\lambda=\varepsilon_p/t$. Notice that $\varepsilon_p$ and $\lambda$, unlike $g$, are adiabatic parameters in the sense that they do not depend of the lattice mass. The phonon energy $\omega_0$ is taken throughout this work as the unit of energy.

\subsection{Electron properties}

In the present work different dynamical correlation functions, which describe properties of the system in the dilute limit, are investigated at frequencies below the phonon threshold. This subsection serves to outline the relevant notation while examining some general aspects of electron correlations. The next subsection, on the other hand, deals with lattice correlations.

The electron propagator, here defined by

\[G_K(t)=-i\langle0|\hat T\;[c _K(t)c_K^\dagger]|0\rangle\;,\]

\noindent with $c_K^\dagger=1/\sqrt N\sum_je^{iKj}c_j^\dagger$ and $\hat T$ the time ordering operator, is evaluated for the ground state of the system $|0\rangle$ in which no electrons are present. As there are no hole excitations for such a state, the spectral function that corresponds to the Fourier transform of $G_K(t)$, $A_K(\omega)=-\pi^{-1}\mbox{Im}G_K(\omega)$, describes only electrons above the polaron ground state energy $E_0$. If $A_K(\omega)$ is decomposed into two parts with respect to the phonon threshold,

\[A_K(\omega)=A^<_K(E_0\leq\omega<\omega_0+E_0)+
A^>_K(\omega\geq\omega_0+E_0)\;,\]

\noindent the low-frequency coherent part of the spectrum can be expressed as

\begin{equation}
A^<_K(\omega)=\sum_i^{\omega<E_0+\omega_0}
A_K^{(i)}\;\;\delta(\omega-E_K^{(i)})\;,\label{A_Kw}
\end{equation}

\noindent with

\[A_K^{(i)}=|\langle\Psi_K^{(i)}|c^{\dagger}_K\rangle|^2\;,\]

\noindent where $|c^{\dagger}_K\rangle$ represents the free electron state without phonons, $E_K^{(i)}$ is the energy and $|\Psi_K^{(i)}\rangle$ is the wave function of the translationally invariant polaron states, and $K$ and $i$ are used to denote the momentum and the number of the polaron band, respectively. At zero temperature, for energies below the phonon threshold $\omega<E_0+\omega_0$, nonelastic scattering is forbidden by energy conservation, which makes the imaginary part of the exact electron (and phonon) self-energy arbitrarily small. For this reason, the incoherent contributions can be neglected in Eq.~(\ref{A_Kw}), and $A^<_K(\omega)$ may indeed be regarded as a signature of the coherent polaron spectrum, but only below the phonon threshold.
 
\subsection{Lattice properties\label{SecLP}}

The phonon propagator used here (not to be confused with the displacement-displacement correlation function) is defined as

\begin{equation}
D_Q(t)=-i\langle\Psi_0|\hat T\;[b _Q(t)b_Q^\dagger]|\Psi_0\rangle\;,\label{bdb}
\end{equation}

\noindent where $|\Psi_0\rangle $ is the ground state of the electron-phonon system. The phonon spectral function $F_Q(\omega)=-\pi^{-1}\mbox{Im}D_Q(\omega)$ is even in $Q$ and generally asymmetric in $\omega$. $F_Q(\omega)$ may be used to spectrally resolve the boson commutation relation

\begin{equation}
\int_0^\infty [F_Q(\omega)-F_Q(-\omega)]\;d\omega=1\;.\label{bsum}
\end{equation}

\noindent The mean number of phonons with the wave vector $Q$ present in the ground state is given by

\begin{equation}
\bar N^{ph}_Q=\langle\Psi_0|b^\dagger_Qb_Q|\Psi_0\rangle
=\int^0_{-\infty} F_Q(\omega)\;d\omega\;.\label{sumrule}
\end{equation}

\noindent The spectral function corresponding to the displacement-displacement correlation function,

\begin{equation}
D^{xx}_Q(t)=-i\langle\Psi_0|\hat T\;
[\hat A_Q(t)\hat A_{-Q}]|\Psi_0\rangle\;,\label{Dxx}
\end{equation}

\noindent with $\hat A_Q=b^\dagger_{-Q}+b_Q$, is written as 

\begin{equation}
P^{xx}_Q(\omega)=F_Q(\omega)+F_Q(-\omega)+2H_Q(\omega)\;,\label{Pxx}
\end{equation}

\noindent where $H_Q(\omega)$, and thus $P^{xx}_Q(\omega)$, is an even function in $Q$ and $\omega$. $H_Q(\omega)$ represents the spectral function related to the correlation function

\[D^{bb}_Q(t)=-i/2\;\langle\Psi_0|\hat T\;[b_Q(t)b_{-Q}
+b_{-Q}^\dagger(t)b_Q^\dagger]|\Psi_0\rangle\;.\]

\noindent It is straightforward to express the spectral functions for the displacement-momentum and momentum-momentum correlation functions in terms of $F_Q(\omega)$ and $H_Q(\omega)$. For the $g=0$ ground state with no phonons, one obtains $F_Q(\omega<0)=0$ and $H_Q(\omega)=0$.

The properties of the general expressions for $F_Q(\omega)$ and $H_Q(\omega)$ can be further specified for the polaron problem. Just as in the electron case, the polaronic correlations manifest themselves in the formation of coherent polaron bands below the phonon threshold. Accordingly, the low-frequency coherent part of the spectral function $F_Q(\omega)$ can be expressed in terms of translationally invariant polaron states,

\begin{eqnarray}
F_Q(\omega)&=&F^<_Q(|\omega|<\omega_0)+
F^>_Q(|\omega|\geq\omega_0)\nonumber\\
F^<_Q(\omega)&=&\sum_i^{|\omega|<\omega_0}
F_{\pm,Q}^{(i)}\;\;\delta(\omega\mp E_Q^{(i)}\pm E_0)\label{F_Q}\;, 
\end{eqnarray}

\noindent with

\[F_{+,Q}^{(i)}=|\langle\Psi_Q^{(i)}|b^{\dagger}_Q|\Psi_0\rangle|^2\;,\;\;\;
F_{-,Q}^{(i)}= |\langle\Psi_{-Q}^{(i)}|b_Q|\Psi_0\rangle|^2\;.\]

\noindent Similarly, the coherent part $|\omega|<\omega_0$ of $H_Q(\omega)$ is given by

\begin{eqnarray}
H_Q^<(\omega)&=&\sum_i^{|\omega|<\omega_0}H_Q^{(i)}\;\;
\delta(\omega\mp E_Q^{(i)}\pm E_0)\label{H_Q}\\
H_Q^{(i)}&=&1/2\;\langle\Psi_0|b_{Q}|\Psi_Q^{(i)}\rangle
\langle\Psi_Q^{(i)}|b_{-Q}|\Psi_0\rangle+\;\mbox{h.c.}\;\;\;.\nonumber 
\end{eqnarray}

\noindent In the absence of electrons, the phonon propagator is purely local, having a trivial temporal dependence, i.e., $F_Q(\omega)=\delta(\omega-\omega_0)$. The phonon spectral weight for $|\omega|<\omega_0$ in Eqs. (\ref{F_Q}) and (\ref{H_Q}) is therefore entirely due to the coupling with the electron, $F^<_Q(\omega), H^<_Q(\omega)\sim1/N$.

In the long-wave limit $Q\rightarrow0$ the phonon spectra in Eqs. (\ref{F_Q}) and (\ref{H_Q}) are smooth functions of $Q$, including $Q=0$. Otherwise, one would find that eigenstates below the phonon threshold involve phonons at infinite distance from the electron. Such phonon excitations uncorrelated to the electron are forbidden by the energy constraint. In Eq.~(\ref{HolHam}) the phonon $Q=0$ mode couples only to the total number of electrons $\sum_kc_k^{\dagger}c_k$, which is a constant of motion. The homogenous $Q=0$ part of the Hamiltonian~(\ref{HolHam}) can therefore be separated out.\cite{Feinberg} Its eigenstates correspond to a displaced harmonic oscillator of frequency $\omega_0$. Assuming one electron in the system, for $|\omega|<\omega_0$ in Eqs.~(\ref{F_Q}) and (\ref{H_Q}) it is straightforward to obtain

\begin{equation}
F_{+,Q=0}^{(i)}=F_{-,Q=0}^{(i)}=H_{Q=0}^{(i)}=
\delta_{i,0}\;\varepsilon_p/N\omega_0\;.\label{Q0}
\end{equation}

The reference case, which illustrates well the fundamental difference between the role of the lowest $i=0$ and the excited $i>0$ polaron bands in the lattice correlation functions, is the limit of self-trapped small polarons ($\varepsilon_p\gg\omega_0$, $\lambda\gg1$). Namely, in this limit the polaron spectrum below the phonon threshold exhibits a few well-separated, narrow polaron bands. The lowest ($i=0$) band contribution to the lattice correlations describes the heavy polaronic lattice distortion, while the excited ($i>0$) bands correspond to the local harmonic modes softened below $\omega_0$, and carried along by the moving polaron. The phonon spectral weight related to the lowest ($i=0$) band involves many phonons, which for small polarons, are equally distributed over all $Q$'s,

\begin{equation}
F_{+,Q}^{(0)}\approx F_{-,Q}^{(0)}\approx H^{(0)}_Q\approx
\bar N_{tot}/N\;,\label{STP}
\end{equation}

\noindent with $\bar N_{tot}\approx\varepsilon_p/\omega_0\gg1$ representing the mean total number of phonons in the ground state,

\begin{equation}
\bar N_{tot}=\sum_Q\langle\Psi_0|b^\dagger_Qb_Q|\Psi_0\rangle \;.\label{Ntot}
\end{equation}

\noindent $\bar N_{tot}$ is large here because of the large lattice distortion that accompanies the formation of the self-trapped polaron. For $|\omega|\leq W^{(0)}$, where $W^{(0)}$ is the bandwidth of the lowest polaron band, one finds from Eqs. (\ref{Q0}) and (\ref{STP}) that the local lattice distortion of the self-trapped small polaron behaves classically for any~$Q$,
 
\begin{equation}
F_{+,Q}^{(0)}-F_{-,Q}^{(0)}\ll1/N\;,\label{class}
\end{equation}

\noindent i.e., as if the boson operators in Eq.~(\ref{bdb}) commute. Hence, $F_Q(\omega)$ as an even function of $\omega$ for $\omega$ small describes the classical nature of the small polaron self-trapping. This represents the generalization of well-known results for large adiabatic polarons\cite{Landau} to the small-polaron case. Furthermore, the distortion moves along the lattice on a long time scale, which is determined by the bandwidth of the lowest polaron band $W^{(0)}$, $W^{(0)}\ll\omega_0$. Such motion is characterized by a small lattice kinetic energy, i.e.,  

\[F_{+,Q}^{(0)}+F_{-,Q}^{(0)}-2H_Q^{(0)}\ll 1/N\;,\]

\noindent where the left-hand side corresponds to the spectral function of the momentum-momentum correlation function.

The total contribution of the $i$th polaron band to the phonon spectral weight can be analyzed in terms of

\begin{equation}
F_+^{(i)}=\sum_Q F_{+,Q}^{(i)}\;,\;\;\;
F_-^{(i)}=\sum_Q F_{-,Q}^{(i)}\;,\;\;\;
H^{(i)}=\sum_Q H_Q^{(i)}\;.\label{ThreeT}
\end{equation}

\noindent Unlike for the lowest ($i=0$) band, the contributions (\ref{ThreeT}) for the excited ($i>0$) bands are independent of parameters; $F_+^{(i)}\approx1$ and $F_-^{(i)}\approx H^{(i)}\approx 0$. The contribution of the excited bands appears only in $F_Q(\omega)$ at the $\omega>0$ side, which is characteristic for harmonic modes that are carried here by the moving polaron, again in generalization of the results obtained in the context of large adiabatic polarons \cite{Melnikov} to the small polaron case.

\subsection{ET method}

The polaron states below the phonon threshold, which define the low-frequency electron and lattice correlations discussed in last two subsections, can be calculated practically exactly by applying the exact-translational (ET) method. This method, introduced in Ref. \onlinecite{Bonca1}, for small and moderate values of the adiabatic parameter $t/\omega_0$ gives the lowest polaron band for the infinite lattice problem with the highest accuracy.\cite{Ku} In addition, the ET method as implemented in Ref. \onlinecite{Barisic2} reveals, with similar accuracy, the excited polaron bands below the phonon threshold. It should be mentioned that the current work provides only a brief overview of the ET method, whereas detailed discussions can be found in Refs.~\onlinecite{Barisic2,Bonca1,Ku,Barisic1}.

The ET method resolves the spectrum below the phonon threshold in terms of the basis set 

\begin{equation}
|n\rangle_K=\frac{1}{\sqrt N}\sum_je^{iKja}c_j^\dagger|
n_0, n_{-1}, n_1,...,n_m,...\rangle_j\;,\label{TBStates}
\end{equation}

\noindent where $n_m$ defines the number of phonons at a given distance $m$ from the electron. On the left hand side of Eq.~(\ref{TBStates}) $n$ is used as an
abbreviation for the whole set of quantum numbers $n_m$.  The translational
symmetry of the problem (\ref{HolHam}) is exploited by the construction of the
states (\ref{TBStates}), which are the eigenstates of the total momentum. In
calculations, the basis set (\ref{TBStates}) is kept finite by restricting the
number of phonons and their distance from the electron. Such
truncation of the basis is justified for the lowest part of the spectrum below
the phonon threshold, which is characterized by the finite electron-phonon
correlation length $d$. Indeed, in the weak coupling regime, $d$ is nonadiabatic, independent of $g$, and determined by the ratio $t/\omega_0$. For strong couplings $d$ is independent of the lattice mass; $d\sim\lambda^{-1}$.

In the representation (\ref{TBStates}), the Hamiltonian is given by a complex Hermitian matrix. However, with the appropriate transformation of the basis (\ref{TBStates}), similar to the one proposed for the antiferromagnetic chain problem in Ref.~\onlinecite{Takahashi},

\begin{eqnarray}
|n_+\rangle_K&=&(|n\rangle_K+|\tilde n\rangle_K)/\sqrt{2}\label{plus}\\
|n_-\rangle_K&=&(|n\rangle_K-|\tilde n\rangle_K)/i\sqrt{2}\label{minus}\;,
\end{eqnarray}

\noindent the Hamiltonian takes the form of a real symmetric matrix. With respect to the previous implementations, this represents a further improvement of the ET method due to the considerable reduction in computational effort associated with it. In Eqs. (\ref{plus}) and (\ref{minus}) $|\tilde n\rangle_K$ is obtained from $|n\rangle_K$ when the phonons to the left and to the right from the electron in Eq. (\ref{TBStates}) are interchanged ($n_m\rightarrow n_{-m}$). If the two states $|n\rangle_K$ and $|\tilde n\rangle_K$ are the same, Eq. (\ref{plus}) should be replaced by the identity, $|n_+\rangle_K=|n\rangle_K$, while $|n_-\rangle_K=0$.

The transformation given by Eqs. (\ref{plus}) and (\ref{minus}) makes use of the point symmetry of the Hamiltonian (\ref{HolHam}), which is invariant under the space inversion. Because of this symmetry, the two subspaces (\ref{plus}) and (\ref{minus}) are not coupled by the Hamiltonian (\ref{HolHam}) for $K=0$ and $K=\pi$. Consequently, for $K=0$ and $K=\pi$ the eigenstates have well defined parity properties: those belonging to the subspace (\ref{plus}) are even, while those belonging to the subspace (\ref{minus}) are odd under the space inversion. Only the two polaron states of opposite parity can cross, as discussed in more details in Sec. \ref{Sec050}.

\section{Nonadiabatic limit}

It is instructive to start the analysis of polaron properties by inspecting the weak- and strong-coupling limits. In this way it is easier to understand the polaron formation in the regime for which neither of the parameters in Eq. (\ref{HolHam}) can be treated as a perturbation. In Sec. \ref{SecWCPT}, the ET results are compared with the weak-coupling perturbation theory (WCPT), while Sec. \ref{SecSCPT} brings the discussion of the strong-coupling perturbation theory (SCPT). In the lowest order, the polaron dispersion is described by nonadiabatic corrections for both these perturbative expansions. The intermediate regime of parameters ($t\sim g\sim\omega_0$), for which the polaron dispersion involves a complex mixture of adiabatic and nonadiabatic effects, is investigated in Sec. \ref{Sec050}. In this latter case one has to rely on numerical results.

\subsection{Weak coupling perturbation theory (WCPT)\label{SecWCPT}}

In the weak coupling limit the deformation field is treated as a perturbation to the free electron state. By applying the \RS\ perturbation theory the energy of the polaron states in the lowest band is, to the leading order in $g$, obtained as 

\begin{equation}
\tilde\xi_k=\xi_k-g^2/\sqrt{(\omega_0-\xi_k)(\omega_0+4t-\xi_k)}
\;,\label{WCPTE}
\end{equation}

\noindent with $\xi_k=2t(1-\cos(k))$. It is assumed in Eq. (\ref{WCPTE}) that the polaron binding energy is small, i.e., $g<\omega_0$ for $t\lesssim\omega_0$ and $g/\omega_0<(t/\omega_0)^{-1/4}$ for $t\gtrsim\omega_0$.\cite{Barisic3} This implies that the lattice distortion for any lattice site is less than the amplitude of the zero-point motion. Accordingly, one finds that, in the weak-coupling limit, the local polaronic correlations and the polaron dispersion are governed by nonadiabatic processes. In the absence of local adiabatic correlations, the spectrum below the phonon threshold consists of only one coherent band.

Up to the second order in $g$, the phonon propagator~(\ref{bdb}) can be evaluated within the diagrammatic perturbation theory by considering the one-loop bubble diagram for the irreducible phonon self-energy. To the leading order in $g$, one obtains a phonon spectral function $F_q(\omega)$ that, beside the phonon branch, exhibits two additional branches at $\pm\xi_q$ defined by the electron dispersion.\cite{Barisic3} For the latter case the spectral density is given by

\begin{equation}
F_q(\omega=\pm\xi_q)=\frac{g^2}{N}\frac{1}{(\omega_0\mp\xi_q)^2}\;.\label{Fq}
\end{equation}

\noindent The remaining spectral density belongs to the phonon-like branch at $\omega=\omega_0$,

\[F_q(\omega=\omega_0)=1+F_q(\omega=-\xi_q)-F_q(\omega=\xi_q)\;.\]

\noindent The last expression, derived here perturbatively, satisfies the boson sum-rule (\ref{bsum}). $F_q(\omega=\xi_q)-F_q(\omega=-\xi_q)$ defines the transfer of the phonon spectral weight from the phonon-like branch $\omega=\omega_0$ to the electron-like branch $\omega=\xi_q$. The additional spectral weight with respect to the free lattice is given by $2F_q(\omega=-\xi_q)$. This is twice the mean number of phonons in the ground state~(\ref{sumrule}).

When $F_q(\omega)$ is known, $H_q(\omega)$ can be determined by calculating the displacement-displacement correlation function (\ref{Dxx}). Up to the second order in $g$ the spectral function $H_q(\omega)$ takes the form

\[H_q(\omega)=\frac{g^2}{N}\frac{1}{\omega_0^2-\xi_q^2}
\left[\delta(\omega\pm\xi_q)-\frac{\xi_q}{\omega_0}\;\delta(\omega\pm\omega_0)\right]\;.\]

\noindent The integrated spectral density $H_q=\int H_q(\omega)\;d\omega$ is related to the static correlation function

\begin{equation}
H_q=\langle\Psi_0|[b_qb_{-q}+b_q^\dagger b_{-q}^\dagger]|\Psi_0\rangle/2=\frac{g^2}{N}
\frac{1}{\omega_0 (\omega_0+\xi_q)}\;.\label{Mbb}
\end{equation}

\noindent Notice that by considering only the first order \RS\ expansion of the polaron wave function $|\Psi_0\rangle$ one obtains $H_q=0$. It is the second-order correction to $|\Psi_0\rangle$ that defines the lowest $g^2$ term in the expansion of~$H_q$. On the other hand, this term is taken into account by the lowest-order diagrammatic expansion of the displacement-displacement correlation function.\cite{Barisic3}

Since the polaron dynamics in the weak-coupling regime is nonadiabatic, the static spatial correlations depend on the lattice mass. In particular, the $q$-dependence of Eqs. (\ref{sumrule}) and (\ref{Mbb}) is governed by the ratio $t/\omega_0$. The electron-phonon correlation length $d$, can be defined as\cite{Romero5}

\begin{equation}
d^2=\sum_{r,n}r^2\;\langle\Psi_0|c^\dagger_nc_n\;
b^\dagger_{n+r}b_{n+r}|\Psi_0\rangle\;/\;\bar N_{tot}\;,
\label{sigma}
\end{equation}

\noindent although other equivalent definitions are possible.\cite{Barisic3,Cataudella} With $\bar N_{tot}$ given by Eq. (\ref{Ntot}), one obtains $d\sim\sqrt{t/\omega_0}$ for $t>\omega_0$ and $d\sim t/\omega_0$ for $t<\omega_0$. Irrespective of $d$, the second-order perturbative results in the weak-coupling limit are very close to the ET findings. This is illustrated by Table~\ref{Ta2}, with $d=0.14$ and $d=2.15$ for $t/\omega_0=0.125$ and $t/\omega_0=10$, respectively.

\begin{table}[hbt]
\begin{center}
\centering{
\begin{tabular}{r|c|c|c|c}
& \multicolumn{2}{c|}{$t=0.125,\;g=0.2$}
& \multicolumn{2}{c}{$t=10,\;g=1$}\\\hline
WCPT & $\bar N_{tot}=0.0272$ & $\bar H=0.0326$
& $\bar N_{tot}=0.08$ & $\bar H=0.156$\\\hline 
ET & $\bar N_{tot}=0.0275$ & $\bar H=0.0328$
& $\bar N_{tot}=0.083$ & $\bar H=0.159$\\\end{tabular}
}\end{center}
\caption{The second-order WCPT is compared to the ET method by means of $\bar N_{tot}$ and $\bar H=\sum_qH_q$, where $\bar N_{tot}$ and $H_q$ are defined by Eqs. (\ref{Ntot}) and (\ref{Mbb}), respectively.\label{Ta2}}
\end{table}

\begin{figure}[tb]

\begin{center}{\scalebox{0.3}
{\includegraphics{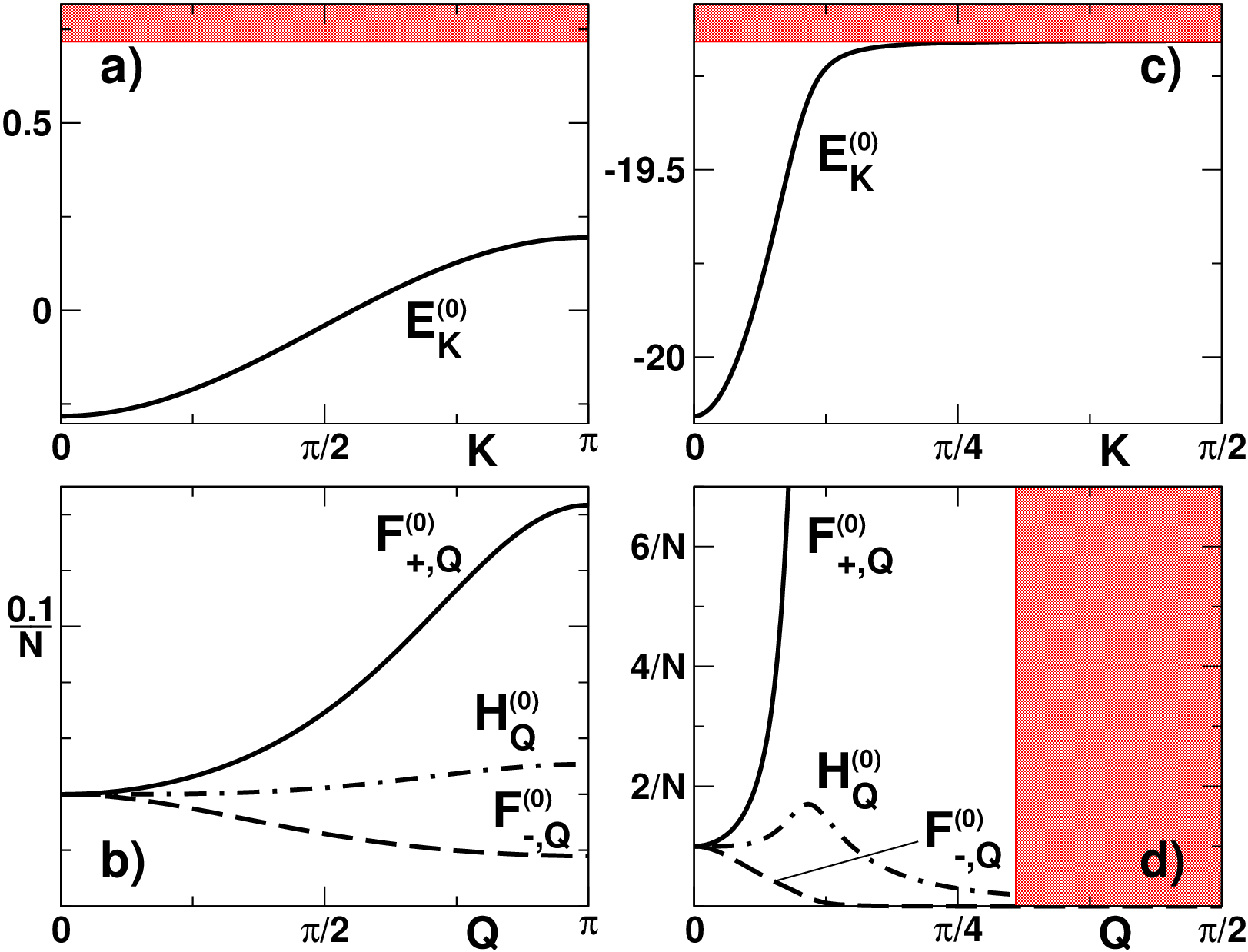}}}
\end{center}

\caption{The lowest coherent polaron band and the lattice correlations in the weak-coupling regime obtained by the ET method. Panels a) and b) are for $t=0.125$ and $g=0.2$, panels c) and d) are for $t=10$, $g=1$. The shaded area in panels a) and c) corresponds to the continuum of states above the phonon threshold that characterizes the exact spectrum of the system. For the ET states that because of a small numerical error are located above the phonon threshold, a part ($K\gtrsim\pi/3$) of panel d) is shaded.\label{fig08}}

\end{figure}

The lattice correlations in the frequency range $|\omega|<\omega_0$ are investigated by the ET method in Fig. \ref{fig08} for the $t<\omega_0$ and $t>\omega_0$ cases, using the same set of parameters as in Table \ref{Ta2}. For $t/\omega_0=0.125$ the coherent polaron band, shown in Fig. \ref{fig08}a, is well-separated from the continuum of states above the phonon threshold (shaded area) for all momenta. The phonon spectra, shown in Fig. \ref{fig08}b, describe a polaronic lattice distortion which exhibits substantial quantum fluctuations, caused by the exchange of momentum between the electron and phonon subsystems. 

As it is well known for weak couplings,\cite{Appel} the nature of the states in the coherent band below the phonon threshold changes as a function of momentum $K$ if $4t>\omega_0$. This change corresponds to a crossover near $K_c$, where $K_c$ denotes the momentum at which the phonon excitation energy is the same as the kinetic energy of the free electron, $\omega_0=\xi_{K_c}$. For $K\lesssim K_c$ the momentum of the system is carried by the polaron (weakly dressed electron), while for $K\gtrsim K_c$ it is carried by the phonon almost uncorrelated to the polaron in the ground $K=0$ state. In the latter case, the wave function of the system involves phonons at arbitrary large distances from the electron. Such long-range correlations, important for $K\gtrsim K_c$ states in the lowest polaron band that are close to the phonon threshold, are not described accurately by the ET approach. Although the exact lowest band remains below the phonon threshold for all $K$, the energy of the ET solutions for $t/\omega_0=10$ in Fig. \ref{fig08}c for $K\gtrsim\pi/3$ ($K_c\approx\pi/10$) is slightly greater than the phonon threshold energy. In order to obtain better results one should use a set of basis states (\ref{TBStates}) that includes phonons at larger distances from the electron than currently considered.
 
The lattice properties are analyzed in Fig.~\ref{fig08}d for $t/\omega_0=10$ by means of ET states satisfying $E_K^{(0)}<E_0+\omega_0$, whereas the shaded area corresponds to the part of the ET spectrum obtained above the phonon threshold ($K\gtrsim\pi/3$). The phonon spectral weight associated with the lowest band diverges in Eq. (\ref{Fq}) for $Q\rightarrow K_c$ as $(\omega_0-\xi_Q)^2$. This singular behavior, replaced by an anticrossing within the degenerate perturbation theory, explains the rapid increase of $F_{+,Q}^{(0)}$ in Fig. \ref{fig08}d. On the contrary, the contributions related to the lattice distortion $F_{-,Q}^{(0)}$ vanish as $K_c$ is approached, with $K_c^{-1}$ of the order of ground-state electron-phonon correlation length (for $4t>\omega_0$). As shown in Fig. \ref{fig08}d, with the exception of the long-wave limit $Q\approx0$, the phonon softening effects are particularly intense, $F_{+,Q}^{(0)}\gg F_{-,Q}^{(0)}$, in the narrow interval $Q<K_c$. This weak-coupling phonon softening, associated with the long-range nonadiabatic correlations, should be distinguished from the softening of the local phonon modes. The latter, obtained for stronger couplings, is due to the short-range adiabatic correlations in the first place.

\subsection{Strong-coupling perturbation theory (SCPT)\label{SecSCPT}}

The small polarons are characterized by a deformation field localized mostly at one lattice site. For small bare electron bandwidths $t<\omega_0$ the electron-phonon correlation length stays short for all couplings, whereas for $t>\omega_0$ it is short when the coupling is sufficiently strong $\varepsilon_p\gg t$ ($\lambda\gg1$). The two regimes, $t\lesssim\omega_0$ and $t\gtrsim\omega_0$, are commonly referred to as the nonadiabatic and the adiabatic regime, respectively, although such identification oversimplifies somewhat the polaron problem. In both cases, the small polarons are frequently discussed using the SCPT. Within the SCPT, it is assumed that the electron dispersion $\varepsilon(k)$ can be treated as a small perturbation of the polaron solution obtained in the atomic limit. The analytical expressions have been derived to the first and second order in $t$ for the infinite lattice,\cite{Marsiglio2} while the third order contributions are analyzed in Ref.~\onlinecite{Firsov2}, but only for the two-site problem. From the low-order expansion in $t$ one is led to believe that the small polarons are necessarily heavy quasiparticles. Indeed, in the adiabatic case one obtains small polarons strongly pinned by the discrete lattice, while in the nonadiabatic case the polaron is heavy because the bare electron hopping is small. It is argued here that, for $t<\omega_0$, the low order SCPT clearly demonstrates the dual character of small-polaron correlations: the nonadiabatic dispersion (translation) and the adiabatic softening associated with the local, adiabatic polaron deformation field. The mixing of nonadiabatic and adiabatic effects becomes particularly important for $t\gtrsim\omega_0$, significantly reducing the small-polaron effective mass. However, the appearance of the light small polarons is beyond the reach of the low-order expansion in $t$ (or $g$ in the context of the WCPT) and therefore analyzed numerically by the ET method in Sec.~\ref{Sec050}. 

In the absence of hybridization, $t=0$, there is no correlation in the lattice dynamics between different sites. At the site with the electron, the deformation field is described by a displaced harmonic oscillator,

\[|\varphi_0\rangle =c^\dagger_j\;e^{-\hat S_j}\;|0\rangle\;,\;\;\;
\hat S_j=-g/\omega_0\;(b_j^\dagger -b_j)\;,\]

\noindent with the (zero-order) binding energy $\varepsilon_0=-\varepsilon_p$. To the first order in $t$ the polaron wave function takes the translationally invariant form,

\begin{equation}
|\varphi_1(k)\rangle=
1/\sqrt N\sum_je^{ikj}c^\dagger_j\;e^{-\hat S_j}|0\rangle\;,\label{SCPT1}
\end{equation}

\noindent corresponding to the well-known Lang-Firsov small-po\-la\-ron band, $\varepsilon_1(k)=-\varepsilon_p+e^{-\gamma}\varepsilon(k)$, where $\gamma=\varepsilon_p/\omega_0$ and $\varepsilon(k)=-2t\;\cos(k) $ is the free electron energy. The dispersion $\varepsilon_1(k)$ is determined purely by nonadiabatic processes: during the hopping to the nearest-neighbor sites the electron detaches itself from the deformation field. The probability of this hopping is reduced with respect to the free electron case by the electron quasiparticle weight

\begin{equation}
|\langle\varphi_1(k)|c_k^\dagger\rangle|^2=e^{-\gamma}\;.\label{LFZk}
\end{equation}

\begin{figure}[tb]

\begin{center}{\scalebox{0.3}
{\includegraphics{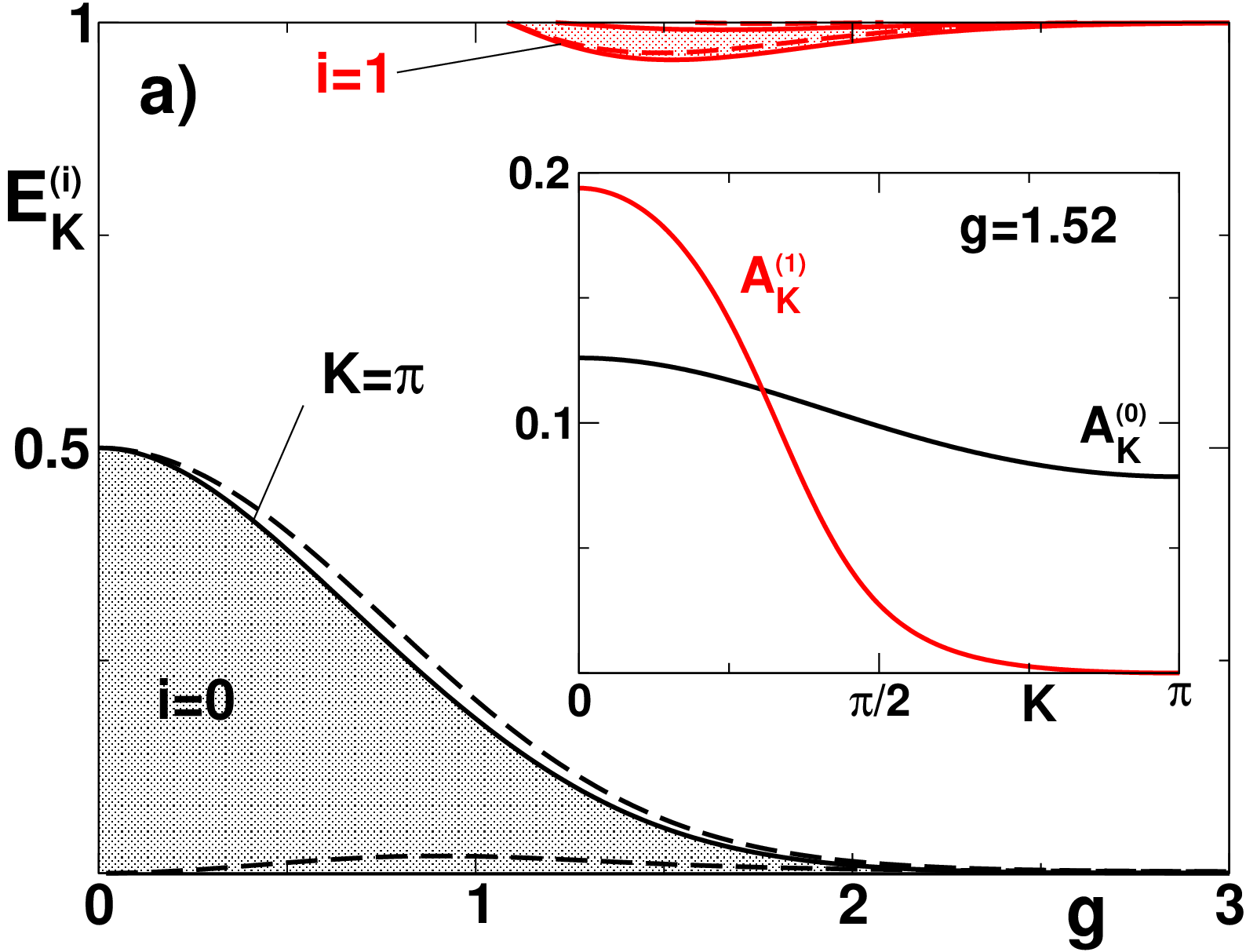}}}
\end{center}

\begin{center}{\scalebox{0.3}
{\includegraphics{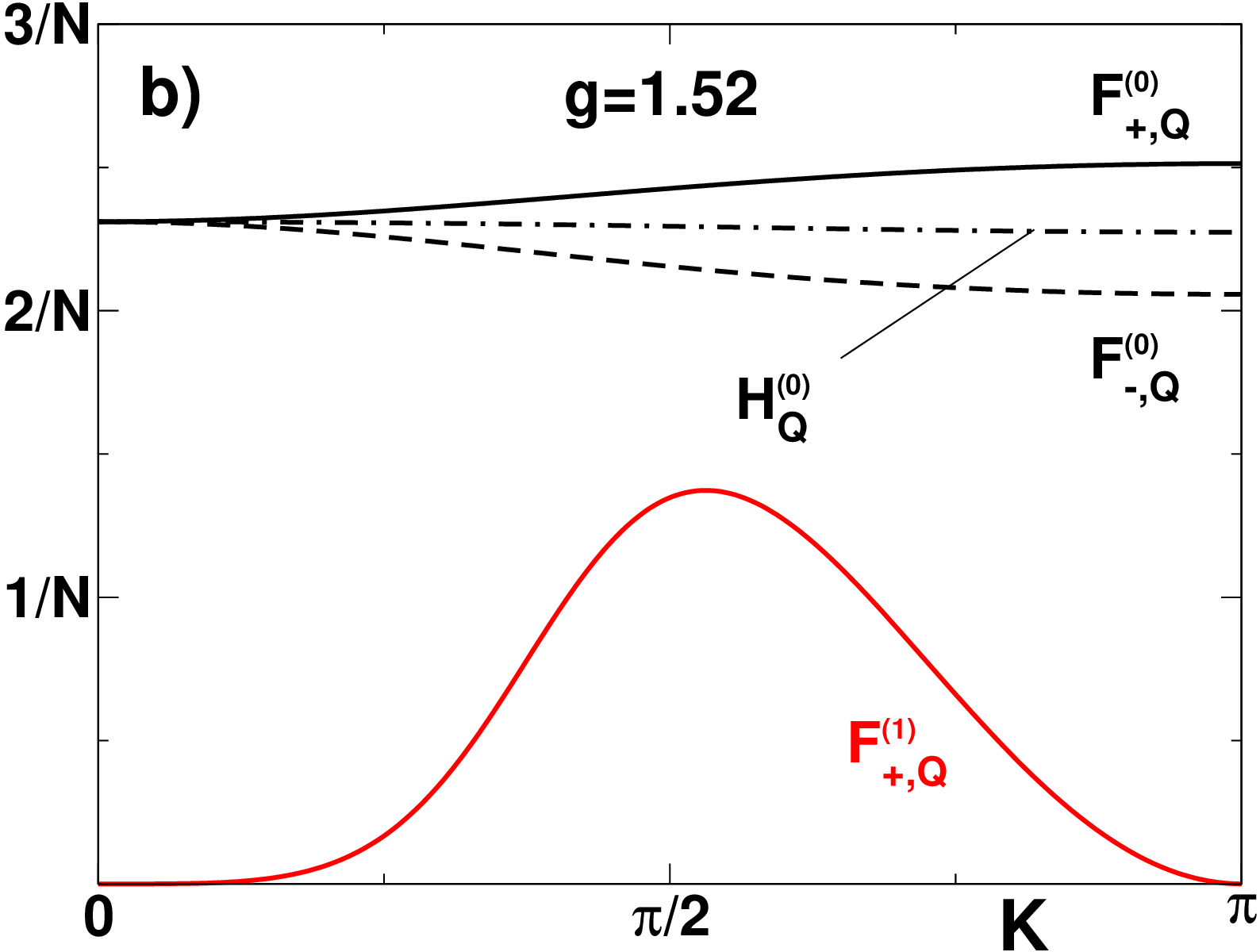}}}
\end{center}

\caption{The two lowest polaron bands below the phonon threshold are plotted in
the upper panel (a) as functions of $g$ for $t=0.125$. The first-order SCPT
results are given by the dashed curves, while the solid curves are the ET
results. The polaron dispersion is taken with respect to the polaron ground
state energy. In the inset of the upper panel the electron spectral function
$A_K^{(i)}$ is shown for $g=1.52$, while the lower panel (b) shows the phonon
spectral functions, $F_{+,Q}^{(i)}$, $F_{-,Q}^{(i)}$ and $H_Q^{(i)}$.\label{fig01}}

\end{figure}

For $t<\omega_0$, the first-order SCPT yields a satisfactory description of the polaron spectrum. This may easily be checked from the weak-coupling side for $t\ll\omega_0$. From the second-order WCPT (\ref{WCPTE}) one obtains then the same polaron energy as from the $\gamma\ll1$ SCPT, $\varepsilon_1(k)\approx-\varepsilon_p+(1-\gamma)\;\varepsilon(k)$. In addition to the lowest band, the SCPT predicts one excited polaron band below the phonon threshold.\cite{Cho} According to the degenerate perturbation theory to the first order in $t$, the first excited band detaches from the bottom of the phonon continuum for $\gamma>1$,\cite{Bonca1}

\[\varepsilon_1^{(1)}(k)=-\varepsilon_p+\omega_0-t\;
e^{-\gamma}\;(x/2+2/x)\;,\]

\noindent with $x=3\gamma-1+\sqrt{(9\gamma-1)(\gamma-1)}$ for $k=0$, while for $k=\pi$ one obtains $x=\gamma$. In Fig. \ref{fig01}a the first-order SCPT low-frequency spectrum (dashed curves), consisting of two bands, when compared for $t=\omega_0/8$ to the exact ET spectrum (solid curves) shows only small deviations. The $i=0$ and $i=1$ bands, shifted by the ground state energy $E_0$, are plotted in terms of band boundaries (the $K=0$ and $K=\pi$ states).

In the inset of Fig. \ref{fig01}a the electron properties are investigated by the ET method for $g/\omega_0=1.52$, while the lattice properties are investigated in Fig. \ref{fig01}b. The choice $g/\omega_0=1.52$ corresponds to the crossover regime between a nearly free-electron behavior, for weak couplings, and an exponentially-narrow-band behavior, for strong-couplings. The exact electron spectral density for the lowest $i=0$ band $A_K^{(0)}$ (the inset of Fig. \ref{fig01}a) as a function of $K$ differs slightly from the first-order SCPT value $e^{-\gamma}\approx0.1$. Similarly, the phonon spectral weight (Fig. \ref{fig01}b) for this band takes values close to Eq. (\ref{STP}), derived for the self-trapped small-polaron limit. In particular, the three contributions of the $i=0$ band to the phonon spectra in Eq. (\ref{ThreeT}) are obtained as $F_+^{(0)}\approx2.43$, $F_-^{(0)}\approx\bar N_{tot}\approx2.18$, $H^{(0)}\approx2.3$. Because $F_-^{(0)}$ and $\bar N_{tot}$ practically coincide, all the spectral weight $F_Q(\omega)$ for $\omega<0$ is assigned to the lowest band. $F_+^{(0)}-F_-^{(0)}\approx0.25$ describes a small departure from the classical behavior (\ref{class}) of the local distortion in Eq.~(\ref{SCPT1}). This signalizes the interference between the polaron motion and the local correlations. For the first excited $i=1$ band, the electron spectral function $A_K^{(1)}$ exhibits a strong $K$-dependence. The phonon spectral weight associated to this excited band indicates that one internal phonon mode of the deformation field is considerably affected by the adiabatic softening; $F^{(1)}_+\approx0.54$ and $F^{(1)}_-\approx H^{(1)}\approx0$. On the strong-coupling side $\lambda\gg1$ of Fig. \ref{fig01}a the excited band asymptotically approaches the phonon threshold, which makes the softening effects particularly small. Such behavior is obtained perturbatively in Ref. \onlinecite{AlexandrovADD}.

For $t>\omega_0$ the SCPT theory converges slowly. The exception is the $\lambda\gg1$ regime, in which the adiabatic contribution to the polaron dynamics is reduced by the very small size of the polaron. Up to the second order in $t$ the polaron energy is\cite{Marsiglio2}

\begin{eqnarray*}
&&\varepsilon_2(k)=-\varepsilon_p+e^{-2\gamma}\varepsilon(0)\;t /\omega_0\;
\left[\Ei\left(2\gamma\right)-\gamma_0-\ln{\left(2\gamma\right)}\right]\\
&&\;+\;e^{-\gamma}\varepsilon(k)+e^{-2\gamma}\varepsilon(2k)\;t/\omega_0\;
\left[\Ei\left(\gamma\right)-\gamma_0-\ln{\left(\gamma\right)}\right]\;,
\end{eqnarray*}

\noindent where $\Ei(\gamma)$ is the exponential integral, and $\gamma_0$ is Euler's constant. In the limit $\gamma \gg1$, the $k$-dependent terms decay exponentially (the first-order nearest-neighbor and the second-order next-nearest-neighbor hopping), while the polaron binding energy becomes strictly adiabatic,

\begin{equation}
\varepsilon_2\approx-\varepsilon_p\;(1+1/\lambda^2)\;.\label{SCPT2}
\end{equation}

\noindent The second-order energy gain in Eq. (\ref{SCPT2}) is related to the spreading of the adiabatic deformation field. Namely, unlike in Eq. (\ref{SCPT1}), where the deformation field develops at the electron site only, the second order wave function 
 
\[|\varphi_2(k)\rangle=1/\sqrt N\sum_{n,m}a_{n,m}^\pm\sum_je^{ikj}c^\dagger_j 
\;(b^\dagger_j)^m(b^\dagger_{j\pm1})^n|0\rangle\]

\noindent involves the electron-phonon correlations at three sites. Apparently, this spreading should also affect the polaron dispersion in higher orders. Indeed, for pinned small polarons it was found by the ET method\cite{Barisic2} that, in the regime $t\sim g$, the lowest band dispersion is described accurately by $e^{-\gamma\;\alpha(\lambda) }\varepsilon(k)$. Here $\alpha(\lambda)\leq1$, which is a function of $\lambda$ only, represents the adiabatic correction to the nonadiabatic first-order hopping.

\section{Adiabatic aspects of small polarons\label{Sec050}}

The polaron self-trapping marks the change from the free-particle-like to the exponentially-large effective mass behavior of the polaron. As shown in Fig.~\ref{fig01}a, for $t<\omega_0$ the polaron dispersion is reasonably well given by the nonadiabatic first-order SCPT for all couplings, including the self-trapping crossover at $g\approx\omega_0$.

For $t\gtrsim\omega_0$, the physical interpretation of the polaron self-trapping constitutes a difficult problem because of the mixing of the nonadiabatic and adiabatic effects. It was first suggested by Eagles\cite{Eagles} that for $t>\omega_0$ the self-trapping may be described by the anticrossing of two qualitatively different, nonadiabatic and adiabatic, polaron states. The first is a large light polaron state, stable on the weak-coupling side. Increasing the coupling, the energy of such large polaron state becomes comparable to the energy of the small heavy polaron, pinned to the discrete lattice. The latter is a stable solution on the strong-coupling side. The sharp, but continuous self-trapping crossover is obtained in terms of the matrix elements involved in the hybridization of the large (nonadiabatic) and the small (adiabatic) state.

\begin{figure}[tpb]

\begin{center}{\scalebox{0.3}
{\includegraphics{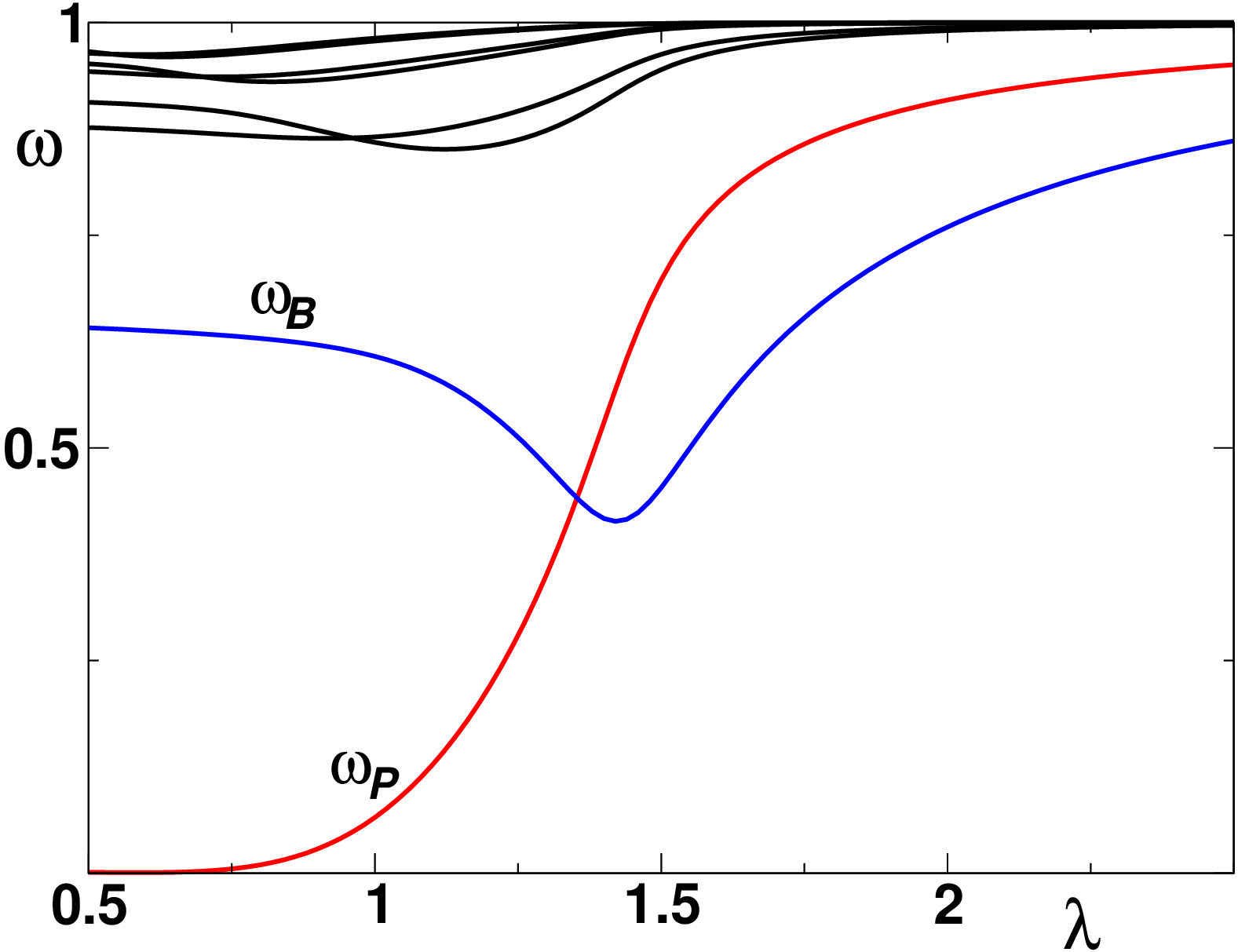}}}
\end{center}

\caption{The 8 lowest normal modes of the polaron deformation, obtained by the harmonic adiabatic approximation, as functions of $\lambda=\varepsilon_p/t$. The pinning and the breather mode are denoted by $\omega_P$ and $\omega_B$, respectively.\label{fig04}}

\begin{center}{\scalebox{0.3}
{\includegraphics{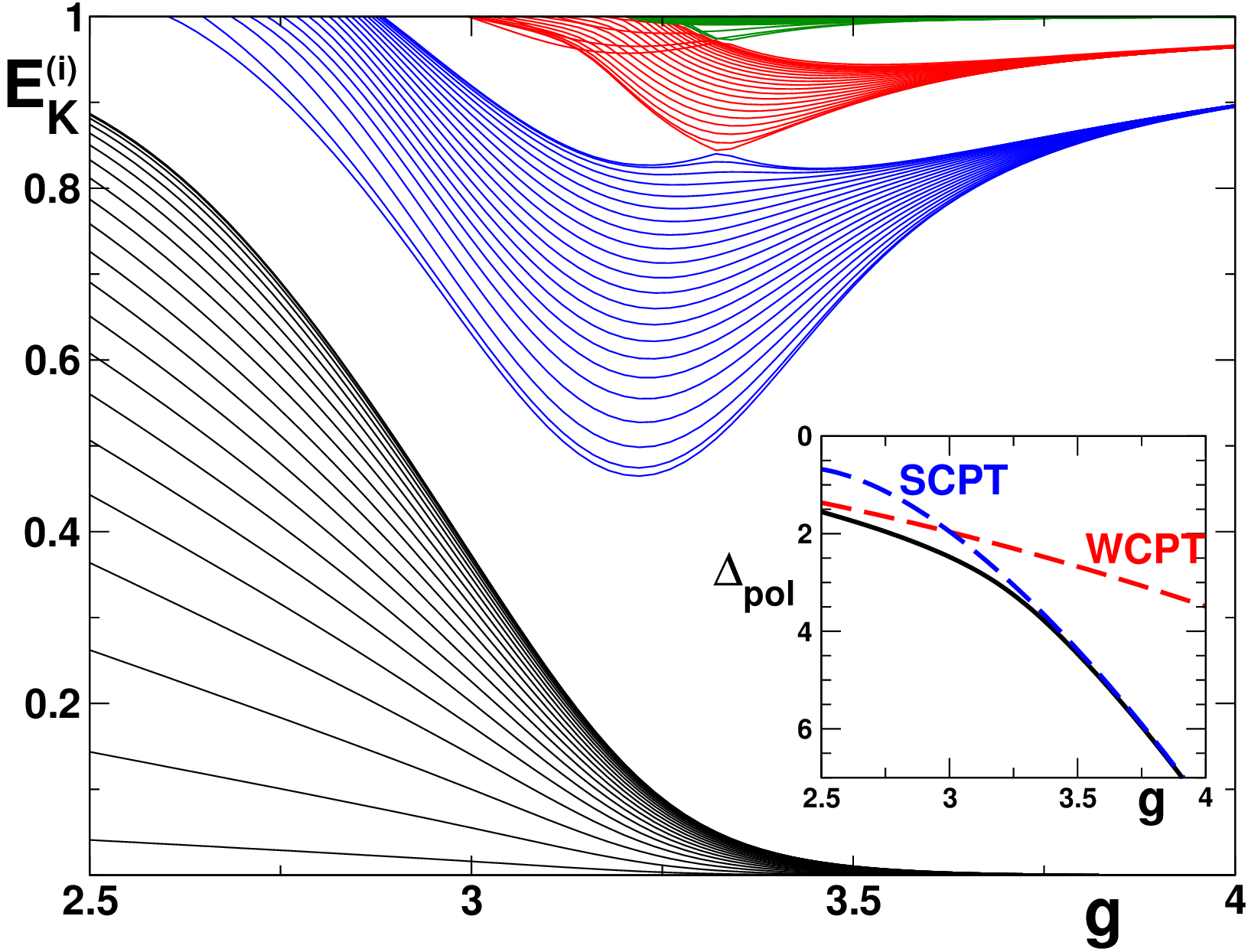}}}
\end{center}

\caption{For $t=5$ the ET polaron spectrum below the phonon threshold, consisting of the lowest and three excited bands, is shown for the crossover regime as a function of $g$. In the inset the exact ET ground-state binding energy $\Delta_{pol}$ is compared to the second-order WCPT and SCPT results.\label{fig05}}

\end{figure}

One might expect (for short-range electron-phonon interactions) that such an anticrossing scenario for $t>\omega_0$ explains correctly the polaron self-trapping for two or more dimensional systems, in which the adiabatic polaron is stable only when it is small,\cite{Kabanov,Kalosakas} exhibiting strong pinning effects. In such circumstances the polaron self-trapping coincides with the nonadiabatic-adiabatic crossover. In 1D, however, the situation is different. In particular, the regime of small adiabatic polarons is reached as $g$ increases by two crossovers in the nature of the ground state, neither of which involves anticrossing among the eigenstates of the Hamiltonian. That is, in the limit $t\gg\omega_0$ the large nonadiabatic weak-coupling polaron crosses into a large adiabatic polaron for $\sqrt{\omega_0/t}\sim\lambda$.\cite{Barisic3} The polaron self-trapping involves a different crossover that occurs for $\lambda\approx1$. For $\sqrt{\omega_0/t}<\lambda\lesssim1$ the adiabatic polaron of the width $d\sim\lambda^{-1}$ behaves practically as a free particle, because the pinning potential is exponentially small.\cite{Kivshar} Within the harmonic adiabatic approximation,\cite{Kalosakas} the crossover from the free-particle-like to the pinned states is reflected in the disappearance of the restoring force near $\lambda\approx1$ for the lowest odd pinning mode, as seen in Fig.~\ref{fig04}. In summary, the polaron self-trapping in 1D can be caused by two different mechanisms. The nonadiabatic one dominates in the $t\ll\omega_0$ limit. It can be described as the simple electron dressing. The other polaron self-trapping mechanism, dominating in the $t\gg\omega_0$ limit, is entirely adiabatic.
 
The particularity of the moderate values of adiabatic ratio $t/\omega_0$ investigated below is that in the crossover regime between weak and strong couplings a mixture of two self-trapping mechanisms, nonadiabatic and adiabatic, occurs. In particular, one finds significant nonadiabatic contributions to the polaron dispersion accompanied with strong adiabatic softening of the two local modes of the deformation field, the latter indicating the weakening of the pinning potential that characterizes the adiabatic polaron motion. The ET spectrum in the crossover regime for $t/\omega_0=5$ is shown in Fig.~\ref{fig05} as a function of $g$. Only the states below the phonon threshold are displayed, shifted by the ground state energy $E_0$. The lowest and the three excited polaron bands are represented each by the $22$ states with different momenta, $K=0,\pi,n\times0.15$, with $n\leq20$. The spectrum in Fig.~\ref{fig05} reveals two important effects of the polaron motion. First, when the bandwidths increase with decreasing $g$, there is a strong hybridization of the first and second excited bands. Second, when $g$ is decreased further all the excited bands shift towards the phonon continuum. This latter effect indicates that the nonadiabatic dynamics prevails in this limit. The fast electron gains more energy by detaching itself from the deformation field, than by waiting to move adiabatically with the lattice distortion to the neighboring site.

\subsection{Strong pinning}

The very narrow polaron bands on the right side of Fig. \ref{fig05} describe self-trapped small-polaron states. The bandwidths $W^{(i)}$ are exponentially reduced by the strong pinning potential due to discreteness of the lattice. The positions of the excited bands in the spectrum are well predicted by the harmonic adiabatic approximation,\cite{Kalosakas} shown in Fig. \ref{fig04}. Because the polarons are small the adiabatic softening is notable only for short-ranged correlations, i.e., for the even (breather) and odd (pinning) modes involving three lattice sites.

The self-trapped polaron represented in terms of Wannier states,

\begin{equation}
|\varphi_n^{(i)}\rangle=1/\sqrt N\sum_K e^{-iKn}|\Psi_K^{(i)}\rangle\;,
\label{Wannier}
\end{equation}

\noindent stays in the quantum state $|\varphi_n^{(i)}\rangle$ centered at the lattice site $n$ over a long period of time, $1/W^{(i)}\gg1/\omega_0$. In such circumstances the spatial ($Q$-dependent) lattice correlations are almost unaffected by the polaron dispersion, i.e.,

\begin{equation}
F_{+,Q}^{(i)}\approx 
|\langle\varphi_n^{(i)}|b^{\dagger}_Q|\varphi_n^{(0)}\rangle|^2\;.\label{Flocal}
\end{equation}

\begin{figure}[tb]

\begin{center}{\scalebox{0.3}
{\includegraphics{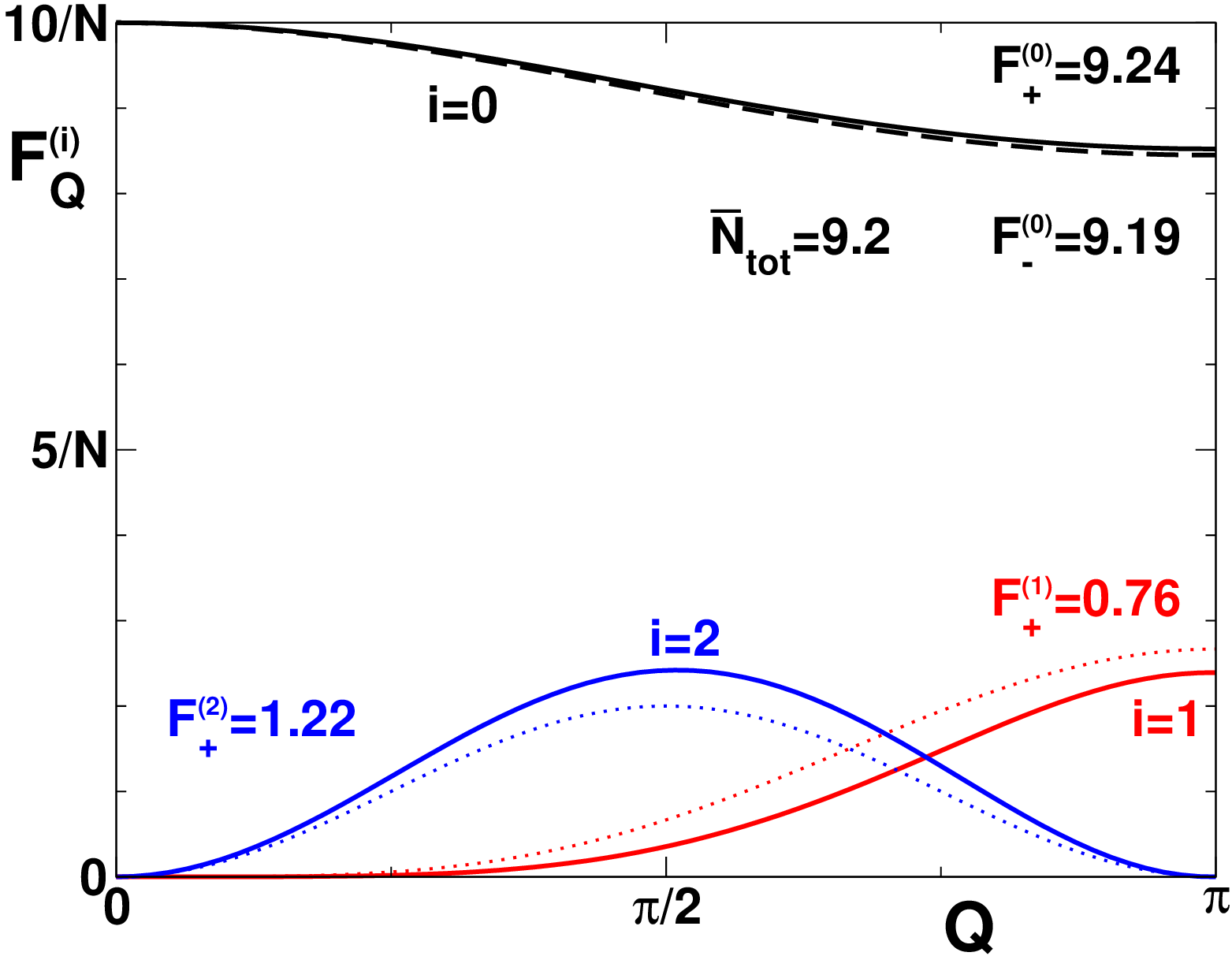}}}
\end{center}

\caption{The ET phonon spectral weight is shown as a function of $Q$ for $\lambda=4$ and $\varepsilon_p/\omega_0=10$. $\bar N_{tot}$ is the mean number of phonons in the ground state (\ref{Ntot}). The thin dotted curves are given by Eqs. (\ref{Breather}) and (\ref{Pinning}). \label{fig02}}

\end{figure}

\noindent As obtained in Eq. (\ref{STP}), in the limit $\lambda\gg1$ the lattice is distorted significantly at one lattice site only, which in the reciprocal space results with the roughly $Q$-independent spectral weight of the lowest polaron band. However, this behavior changes by decreasing $\lambda$ as the distortion spreads to neighboring sites. For $\lambda=4$ and $\varepsilon_p/\omega_0=10$ the phonon spectra are plotted in Fig.~\ref{fig02}. The parameters used correspond to the self-trapped polaron regime, in which $F_{+,Q}^{(0)}$, $F_{+,Q}^{(0)}$, and $H_{Q}^{(0)}$ are practically the same (classical distortion). Hence, the decrease of the $i=0$ spectral weight toward the end of the Brillouin zone, which is seen in Fig. \ref{fig02}, is related solely to the spreading of the local distortion. This spreading is characterized by the adiabatic correlation length $d\sim1/\lambda$ as long as the distortion behaves classically. Increasing the range of the adiabatic correlations (at smaller $\lambda$) only makes the $Q$-dependent feature of the $i=0$ spectral weight observed in Fig. \ref{fig02} more pronounced.

In the $\lambda\gg1$ limit, the two lowest internal soft-phonon modes of the small-polaron deformation field involve three lattice sites, with a displacement pattern $(1,-2,1)$ for the even (breather), and (-1,0, 1) for the odd (pinning) mode.\cite{Kalosakas} The corresponding phonon operators with the appropriate normalization are given by

\[b^{(1)}_n=\frac{b_{n-1}-2b_n+b_{n+1}}{\sqrt{6}}\;,\;\;\;
b^{(2)}_n=\frac{b_{n+1}-b_{n-1}}{\sqrt{2}}\;.\]

\noindent If the adiabatic fluctuations of the electron in Eq. (\ref{Flocal}) are neglected, the contribution of which is of the order $\omega_0/\varepsilon_p$, the phonon spectral density for the breather mode is obtained as

\begin{equation}
F_{+,Q}^{(1)}\approx
|\langle0|b^{(1)}_nb^{\dagger}_Q|0\rangle|^2=
\frac{2}{3N}\;\left(1-\cos{(Q)}\right)^2\;,\label{Breather}
\end{equation}

\noindent and for the pinning mode as

\begin{equation}
F_{+,Q}^{(2)}\approx
|\langle0|b^{(2)}_nb^{\dagger}_Q|0\rangle|^2
=\frac{2}{N}\;\sin^2{(Q)}\;.\label{Pinning}
\end{equation}

\noindent Here, the lattice distortion is not considered explicitly as it does not contribute to $F_{+,Q}^{(1,2)}$. The expressions (\ref{Breather}) and (\ref{Pinning}) explain the spatial correlations obtained by the ET method in Fig. \ref{fig02} for the two lowest $i=1,2$ excited bands of the self-trapped small polaron. Namely, the thin dotted curves, representing Eqs. (\ref{Breather}) and (\ref{Pinning}), compare well to the ET results for $F_{+,Q}^{(1)}$ (breather) and $F_{+,Q}^{(2)}$ (pinning).
 
In contrast to the $i=0,1,3$ bands in Fig. \ref{fig05}, the spectrum of the second $i=2$ excited band is inverted. The bottom and the top of this band are associated to the $K=\pi$ and the $K=0$ state, respectively. Such dispersion is a consequence of the parity properties of the local adiabatic correlations. This is easily understood by analyzing the self-trapped small-polaron limit, in which the hybridization between different bands is negligible. In this case, the Wannier states (\ref{Wannier}) inherit the symmetry of the internal phonon modes of the deformation field. The pinning ($i=2$) mode is odd under the space inversion, which results in a positive nearest-neighbor intraband hopping.

\subsection{Crossover regime\label{Sec052}}

The bandwidths of the self-trapped polaron bands increase by decreasing the coupling due to the weakening of the pinning potential. The pinning potential is more effective for the polaron states at the bottom of the spectrum, explaining why the bandwidths of the two lowest excited bands on the right side of Fig. \ref{fig05} are significantly larger than the bandwidth of the lowest band.
In the crossover regime, corresponding to the central part of Fig. \ref{fig05}, the bandwidths and the positions of the bands in the spectrum are characterized by the same time scale. This means that the local correlations, dominated in the self-trapped polaron limit by the harmonic adiabatic dynamics, in the crossover regime are strongly affected by anharmonic and nonadiabatic contributions introduced by the small-polaron hopping.

\begin{figure}[tb]

\begin{center}{\scalebox{0.3}
{\includegraphics{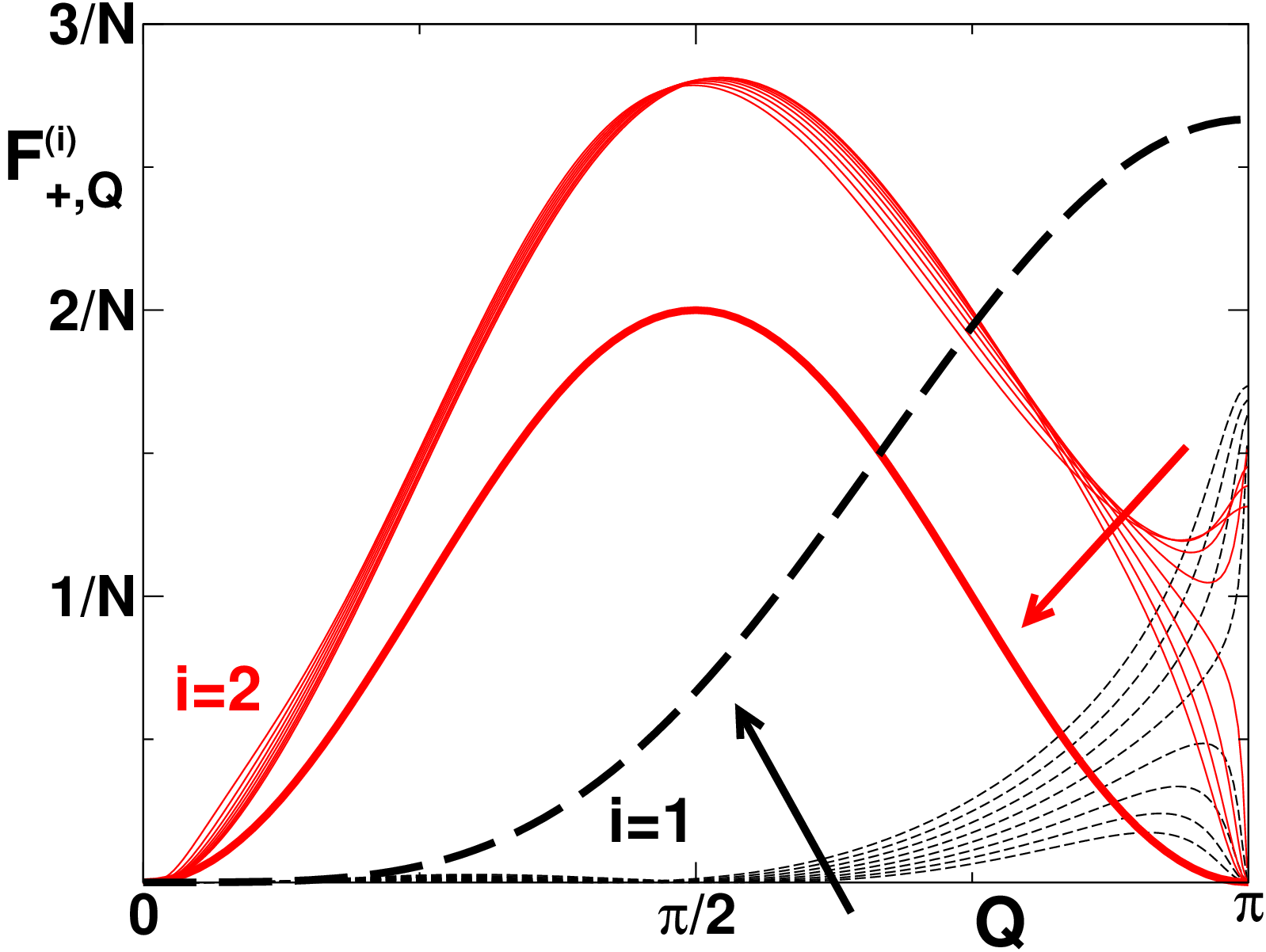}}}
\end{center}

\caption{ The phonon spectral weight $F_{+,Q}^{(i)}$, as a function of $Q$, for the first $i=1$ (dashed curves) and the second $i=2$ (solid curves) excited band. The results are obtained for the set of couplings that corresponds to the narrow interval in which the two $i=1$ and $i=2$ excited bands touch ($2.82\leq g_n\leq2.89$, with $g_{n+1}=g_n+0.01$, $t=2.5$). The arrows point to the direction of the increasing coupling. The thick dashed and solid curves are plotted for the self-trapped small-polaron limit in Eqs. (\ref{Breather}) and (\ref{Pinning}).\label{fig06}}

\end{figure}

In the central part of Fig. \ref{fig05} the polaron spectrum is characterized by the strong hybridization of the first ($i=1$) and the second ($i=2$) excited bands, while the hybridization between states of the second ($i=2$) and third ($i=3$) excited bands is weak. The $i=1$ and $i=2$ bands touch for $g\approx3.32$. This band touching corresponds to the symmetry allowed crossing of the two $K=\pi$ states of opposite parity. The origin of the strong hybridization of the two lowest excited bands lies with the small-polaron hopping. While the motion of the large polaron to the neighboring site corresponds to a small change of the distortion, the small polaron moves by creating the lattice distortion at one site and destroying it at the other. The displacement pattern involved in the nearest-neighbor small-polaron hopping to the right is approximately given by $(0,-1,1)$. It is easy to see that such a pattern is obtained by a linear combination of the breather $(1,-2,1)$ and pinning $(-1,0,1)$ displacement patterns. Thus, for small polarons the breather and the pinning mode are, due to discreteness of the lattice, effectively coupled by the polaron motion. Figure~\ref{fig06} demonstrates the correlation in the behavior of the phonon spectral weight for the $i=1$ and $i=2$ bands in the touching regime. Figure~\ref{fig06} shows that, as the hybridization with the increasing coupling becomes weaker, there is a rapid transfer of the breather-like phonon spectral weight near the end of the Brillouin zone from the $i=2$ band (solid curves) to the $i=1$ band (dashed curves).

\begin{figure}[tb]

\begin{center}{\scalebox{0.4}
{\includegraphics{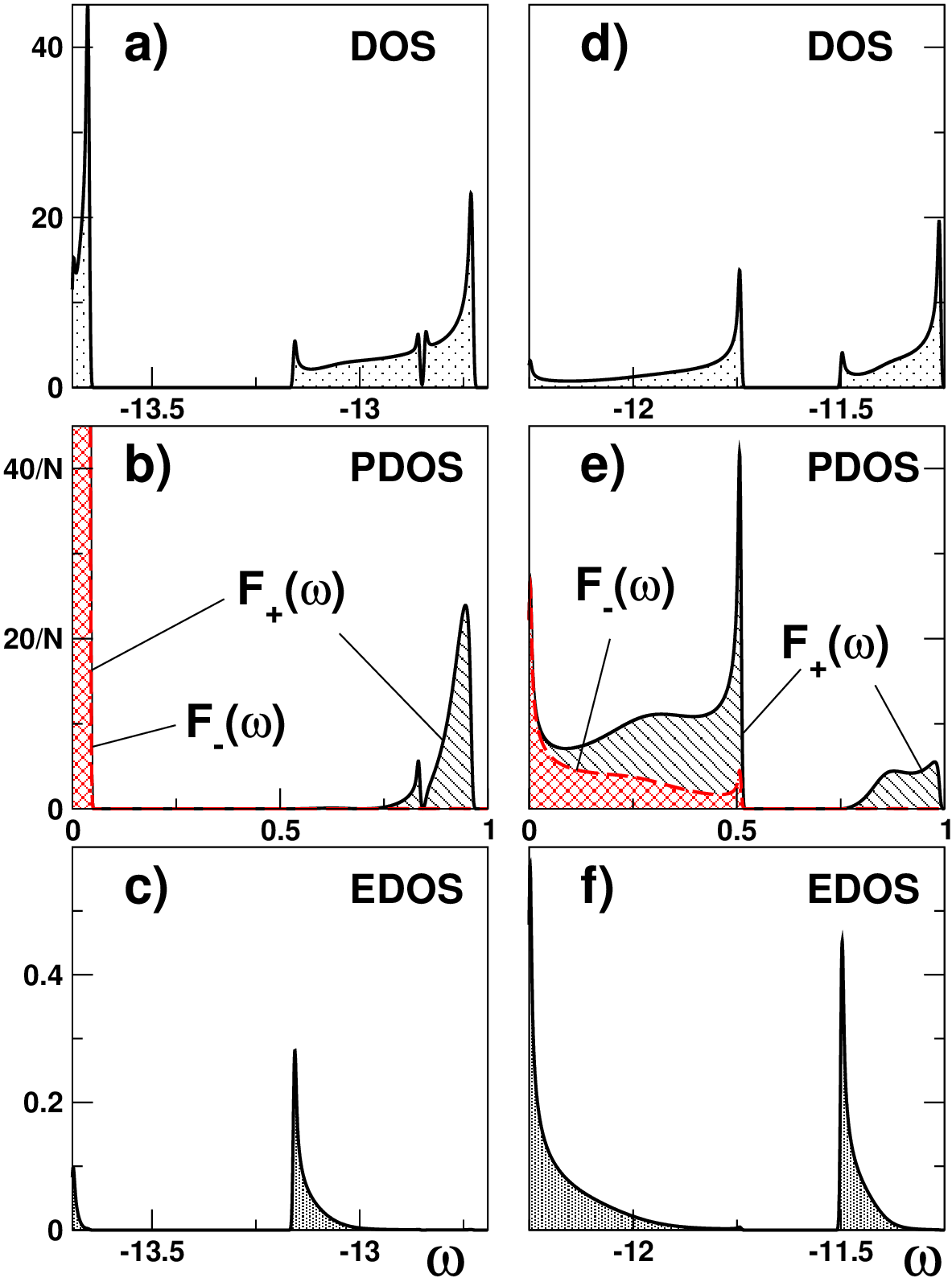}}}
\end{center}

\caption{The polaron (DOS), phonon (PDOS), and electron (EDOS) density of states for frequencies below the phonon threshold. The first and the second column of panels are given for $g=3.35$ and $g=2.9$, respectively, with $t=5$ in both cases. The amplitude of the PDOS exceeds the range of values shown in Fig. \ref{fig10}b.\label{fig10}}

\end{figure}

The polaron density of states (DOS),

\[S(\omega)=1/N\;\sum_KS_K(\omega)\;,\;\;\;
S_K(\omega)=\sum_i\delta(\omega-E_K^{(i)})\;,\]

\noindent the phonon density of states (PDOS), $F(\omega)=1/N\;\sum_QF_Q(\omega)$, and the electron density of states (EDOS), $A(\omega)=1/N\;\sum_KA_K(\omega)$, are investigated in Fig. \ref{fig10} for the crossover regime ($t/\omega_0=5$ as in Fig. (\ref{fig05})). The left column of panels presents the coherent low-frequency features at the strong-coupling side of the crossover regime, $g/\omega_0=3.35$, whereas the weak-coupling side of the crossover regime is analyzed in the right column, $g/\omega_0=2.9$. For the former case the polaron effective mass is large, $m_{pol}/m_{el}\approx100$, while for the latter case it is within the order of magnitude of the electron effective mass, $m_{pol}/m_{el}\approx5$.

The DOS in Fig. \ref{fig10}a contains three distinct features, each corresponding to one of the three lowest polaron bands in Fig. \ref{fig05} (the contribution of the third excited band located in Fig. \ref{fig05} very close to the phonon threshold is not shown in Fig. \ref{fig10}). Three distinct contributions may also be observed in the PDOS of Fig. \ref{fig10}b. The low frequency feature, $\omega\sim W^{(0)}$, in Fig. \ref{fig10}b is given by the narrow, $W^{(0)}\ll\omega_0$, lowest band, for which the phonon spectral weight  behaves approximately classically, $F(\omega)\approx F(-\omega)$. Because of the weak $Q$-dependence of the phonon spectral weight (small-polaron distortion) the PDOS and DOS for the lowest band have similar shape. The considerable flattening of the lowest polaron band toward the end of the Brillouin zone is the reason for the more pronounced singular behavior of the DOS (and the PDOS) at the upper rather than at the lower band edge.

The two lowest excited bands, related to the pinning and breather modes of the deformation field, almost touch in Fig. \ref{fig10}a. Their bandwidths are significantly larger than the bandwidth of the lowest band, $W^{(1)}+W^{(2)}\approx\omega_0/2$. The phonon spectral weight assigned to the excited bands, accruing only for $\omega>0$, exhibits a strong $Q$-dependence. This explains the substantial difference in the shape of the DOS in Fig. \ref{fig10}a and the PDOS in Fig. \ref{fig10}b for frequencies corresponding to the $i=1$ and $i=2$ bands. The hybridization of the breather and pinning modes due to the polaron hopping can be observed in the partial transfer of the phonon spectral weight from the $i=1$ to the $i=2$ band. 

For $g/\omega_0=3.35$ the electron properties are shown in Fig. \ref{fig10}c.
Although the polaron is small and heavy, the EDOS of the lowest band differs essentially from the simple momentum-independent Lang-Firsov formula (\ref{LFZk}). The most significant contributions to the EDOS involve the $i=0$ and $i=1$ states near the bottom of the bands. In fact, one finds that, in the crossover regime, the electron spectral weight for the lowest band is weakly dependent on the momentum only for small bare electron bandwidths (Fig. \ref{fig01}). The contribution of the $i=2$ band to EDOS in Fig. \ref{fig10}c is very small because of its parity properties. While the free electron state belongs to the subspace (\ref{plus}), the main contribution to the polaron states of the $i=2$ band is given by the states belonging to the subspace (\ref{minus}). In particular, the $K=0$ and $K=\pi$ states of the $i=2$ band are odd under space inversion, with no overlap whatsoever with the free electron state of even parity.\cite{Barisic2}

For $g/\omega_0=2.9$ the stable polarons are small and light, exhibiting particularly interesting properties. Figure \ref{fig10}d shows the polaron spectrum below the phonon threshold consisting of two coherent bands. The bandwidth of the lowest band is larger than that of the excited band. The PDOS is shown in Fig. \ref{fig10}e. For the lowest band $F(\omega)$ shows notable asymmetry in $\omega$, which means that the local lattice distortion begins to exhibit significant quantum low-frequency fluctuations associated with the polaron motion. Such quantum behavior is an indication of nonadiabatic effects. The adiabatic effects are manifested as a softening of the internal phonon mode of the deformation field. This softening is represented in Fig. \ref{fig10}e by the excited band contribution, found only for $\omega>0$. The EDOS in Fig. \ref{fig10}f is, in a manner similar to that seen in Fig. \ref{fig10}c, characterized by two pronounced singularities at the bottom of the polaron bands. For $g/\omega_0=2.9$ (as opposed to $g/\omega_0=3.35$) the polaron states are lighter for the lowest band than they are for the excited band. Accordingly, one finds that the lowest band contributes more to the electron spectral weight than the excited one in Fig. \ref{fig10}f. Such a dispersion can be is explained by nonadiabatic dynamics that is energetically more favorable for the polaron motion than the adiabatic one.

\begin{table}[t]
\begin{center}
\centering{
\begin{tabular}{r|c|c|c|c}
& $\Delta_{pol}$ & $d$&$\bar N_{tot}$ 
&$m_{el}/m_{pol}$\\\hline
$g=3.35$ & $\;3.69\;$ & $\;0.45\;$ & $\;7.56\;$
& $\;0.01\;$\\\hline
$g=2.9$ & $2.25$ & $0.93$ & $2.25$
& $0.196$
\end{tabular}
}\end{center}
\caption{Comparison of small-polaron properties for $g=3.35$ and $g=2.9$ ($t=5$).\label{Ta1}}
\end{table}

Table \ref{Ta1} brings some additional insights upon the choice of parameters used in Fig. \ref{fig10}. $\Delta_{pol}$ is the binding energy, the electron-phonon correlation length $d$ is defined by Eq. (\ref{sigma}), $\bar N_{tot}$ is given by Eq. (\ref{Ntot}), and $m_{el}/m_{pol}$ determines the polaron effective mass. By going from $g/\omega_0=3.35$ to $g/\omega_0=2.9$ the static polaron properties ($\Delta_{pol}$, $d$, $N_{tot}$) change rapidly, but far more dramatic is the change in the dispersion. For $g/\omega_0=2.9$ the polaron effective mass is within the order of magnitude of the electron effective mass, the latter being characterized by the large bare electron bandwidth, $t/\omega_0=5$. It is emphasized that for such a large ratio of $t/\omega_0$, the motion of the polaron in the crossover regime involves substantial nonadiabatic and adiabatic effects. In the nonadiabatic case, the hopping occurs by a dephasing between the electron and the lattice distortion. The adiabatic dynamics contributes to the small-polaron dispersion through the softening of the internal phonon mode, which for small polarons, is obtained as a mixture of the breather and pinning modes. Because of this adiabatic softening the effective hopping overlap between the polaron states at neighboring sites is larger than for the unrenormalized lattice vibrations.

\section{Summary and conclusions}

The Holstein polaron problem is analyzed here for the one-dimensional infinite lattice. New important insights into the polaron formation are obtained and the crossover between the light and heavy polaron states is clarified. Using the practically exact polaron states, obtained numerically by the ET method, the spatial and temporal dependence of electron and lattice correlations is examined for the coherent part of the spectrum below the phonon threshold. Special attention is paid to the properties of the phonon propagator and the related correlation functions. To our knowledge, this constitutes the first investigation of such kind in terms of translationally invariant polaron states.

In order to achieve a better understanding of nonadiabatic and adiabatic effects in the polaron dispersion and of the local correlations, the weak- and strong-coupling perturbative results and the results of the harmonic adiabatic approximation are compared to those of the ET method. The polaron dispersion may involve both adiabatic and nonadiabatic processes. In the first case, the electron is described by a localized state that follows adiabatically the distortion as it moves along the lattice. In the second case, the electron hops away from the distortion before self-localizing again. The local adiabatic correlations are responsible for the softening of the internal phonon modes of the polaron. Each of the excited coherent polaron bands found below the phonon threshold corresponds to one of these soft phonon modes. 

The distinction between nonadiabatic and adiabatic contributions provides a qualitative description of the polaron crossover between the self-trapped and free-like-particle behaviors of the electron, in the intermediate regime $t\sim\omega_0\sim g$ in which neither of the parameters can be treated as a small perturbation. For $t\sim\omega_0\sim g$ the crossover regime involves the small polaron states, for which the adiabatic motion along the lattice is governed by a pinning potential due to the discrete lattice. It is argued that the adiabatic contributions to the effective nearest-neighbor hopping integral of such polarons are enhanced by the softening of the two lowest (even and odd) internal modes. In return, the polaron motion effectively couples these two modes, which results in a hybridization of the states in the first- and second-excited polaron bands below the phonon threshold. This hybridization is weak on the strong-coupling side of the crossover regime for which the excited bands are narrow and well-separated. Because the pinning potential affects the low-frequency states more, the lowest band is notably narrower than the first two excited bands. As the coupling is decreased, the bandwidths of all the bands increase progressively because the pinning potential weakens. In addition, the hybridization between the two lowest excited bands becomes strong. The shifting of all the excited bands towards the phonon threshold $\omega_0$ on the weak-coupling side of the crossover regime indicates that the local adiabatic correlations are suppressed by the nonadiabatic contributions to the dispersion. In this regime of competing adiabatic and nonadiabatic effects, one finds that the small polarons are light. The binding energy $\Delta_{pol}$ is a few $\omega_0$, while the spectrum below the phonon threshold consists of two coherent polaron bands. The polaron dynamics becomes fully nonadiabatic in the weak-coupling limit, in which the coherent part of the spectrum below the phonon threshold exhibits only the lowest polaron band. The weak interaction between the electron and the lattice distortion is reflected in the small binding energy ($\Delta_{pol}<\omega_0$).

The lattice correlations change remarkably among different regimes of parameters. The behavior of the phonon spectra is particularly complex in the crossover regime, in which the nonadiabatic and adiabatic contributions mix.  The crossover regime connects two essentially different limiting behaviors obtained for strong and weak couplings, respectively. In the strong-coupling limit (self-trapped polarons) the phonon spectral weight at very low frequencies describes a classical lattice distortion, while for frequencies corresponding to the soft internal phonon modes, the spectral weight is typical for harmonic dynamics. The small $|\omega|$ contribution, defined by the distortion in the ground state, corresponds to the additional spectral weight that is found with respect to the free lattice. The phonon softening involves the transfer of the spectral weight from the unperturbed frequency $\omega_0$ to lower frequencies. As the slow quantum motion of the self-trapped polaron only weakly affects the local dynamics, the $Q$-dependence of the phonon spectral functions describes the local spatial correlations. On the contrary, for weak couplings the polaronic distortions exhibit strong nonadiabatic quantum fluctuations due to the exchange of momentum between the electron and lattice subsystems. In particular, as the lowest polaron band approaches the phonon threshold, significant phonon softening effects are associated with long-range nonadiabatic correlations. The phonon spectral weight which corresponds to this softening can be much larger than that related to the weak-coupling distortion, which is defined by a small number of phonons in the ground state.

\begin{acknowledgments}

Useful discussions with prof. S. Bari\v si\' c are acknowledged. This work was
supported by the Croatian Government under Project No. $0035007$.

\end{acknowledgments}

\end{document}